\documentclass[conference]{IEEEtran}
\IEEEoverridecommandlockouts
\usepackage{cite}
\usepackage{amsmath,amssymb,amsfonts}
\usepackage{algorithmic}
\usepackage{graphicx}
\usepackage{textcomp}
\usepackage{xcolor}
\def\BibTeX{{\rm B\kern-.05em{\sc i\kern-.025em b}\kern-.08em
    T\kern-.1667em\lower.7ex\hbox{E}\kern-.125emX}}
\begin{document}

% \title{Evaluating Variants of wav2vec 2.0 \\ on Affective Vocal Bursts Tasks}

\title{Predicting Affective Vocal Bursts \\ with Finetuned wav2vec 2.0}

% \and
% \and
% \and
% \and
% \and
\author{\IEEEauthorblockN{Bagus Tris Atmaja}
\IEEEauthorblockA{\textit{AIST} \\
Tsukuba, Japan \\
% https://orcid.org/0000-0003-1560-2824
b-atmaja@aist.go.jp
}
% \and
% \IEEEauthorblockN{Zanjabila}
% \IEEEauthorblockA{\textit{ITS} \\
% % \textit{name of organization (of Aff.)}\\
% Surabaya, Indonesia \\
% zanjabilaabil@gmail.com}

\and
\IEEEauthorblockN{Akira Sasou}
\IEEEauthorblockA{\textit{AIST} \\
Tsukuba, Japan \\
a-sasou@aist.go.jp}
}

\maketitle
% focus on finetuned
\begin{abstract}
The studies of predicting affective states from human voices have relied heavily on speech. This study, indeed, explores the recognition of humans' affective state from their vocal burst, a short non-verbal vocalization. Borrowing the idea from the recent success of wav2vec 2.0, we evaluated finetuned wav2vec 2.0 models from different datasets to predict the affective state of the speaker from their vocal burst. The finetuned wav2vec 2.0 models are then trained on the vocal burst data. The results show that the finetuned wav2vec 2.0 models, particularly on an affective speech dataset, outperform the baseline model, which is handcrafted acoustic features. However, there is no large gap between the model finetuned on non-affective speech dataset and affective speech dataset.
\end{abstract}


\section{Introduction}
% Affective computiong
Speech emotion recognition is currently gaining more attention from researchers due to its potential implementation in the market. Instead of speech, affective information may also lay on short vocal bursts (i.e., cry when sad). In contrast to speech emotion recognition which may have difficulties in distinguishing between emotions (for example, the emotion of anger and fear are similar since both of them are expressed by raising the pitch of the voice), different vocal bursts may reflect different affective states more distinctly. A specific pattern of crying may indicate sadness, whereas laughter produces happiness. Given these benefits, analyzing emotions from humans' vocal bursts may improve our understanding of human emotions.

% vocal burst
Vocal bursts, a non-verbal communication, constitute a potential source of information for emotion \cite{Holz2021}. A study by Cowen et al. \cite{Cowen2019} has found that vocal bursts are rich in emotional information that can be conceptualized into 24 emotion categories. A previous study by Scherer \cite{Scherer2003} has proposed a model of vocal communication as Brunswik's lens model from expression (encoding) to perception (representation). There is no exact number of emotion categories resulting from this study. The authors mentioned eight examples of emotion categories with ranges of importance for their design features delimitation (e.g., intensity). However, the research on affective vocal bursts since then has been limited by the lack of available datasets.

% Need to be rewritten from the original APSIPA submission
% dataset
One way to speed up research on affective vocal bursts is to hold workshops and competitions in that area. In \cite{Baird2022,Schuller2022,Baird2022a}, the organizers provided datasets and baseline methods to challenge participants to explore the dataset and surpass the baseline scores. This technical report, in particular, is presented to report the evaluation of wav2vec 2.0 variants finetuned on specific datasets for the ACII 2022 Affective Vocal Bursts Workshop and Competition \cite{Baird2022a}. There are four tasks in the competition: three regression problems for measuring either the intensity of emotion categories or valence and arousal and a classification problem for predicting the type of vocal bursts. We approached all of the four tasks using a similar method while varying the acoustic embeddings.

% contribution
Therefore, the main contribution of this study to affective vocal bursts research is the evaluations of different finetuning methods based on wav2vec 2.0 for affective vocal burst tasks. Five variants of wav2vec 2.0 including the base model with large and robust version -- without finetuning -- are applied on the four affective vocal bursts tasks held in the 2022 ACII Affective Vocal Bursts Workshop and Competition \cite{Baird2022a}. The architecture of deep learning models was similar for all tasks except in the output layer and the loss function (which depends on the task). The results show the improvements in the use of wav2vec 2.0 variants over the baseline method with the conventional acoustic features.
% to be cited: APSIPA draft, ACII draft, Sensors

\begin{table*}[hbt!]
    \caption{Base and finetuned of wav2vec 2.0 and their Hugging Face model; the last two rows used the same model but different sizes.}
    \centering\begin{tabular}{l l l l}
    \hline
    Name & Finetuning dataset & Hugging Face Model ID & Reference\\
    \hline
    w2v2-lr		    & None & facebook/wav2vec2-large-robust & \cite{Hsu2021a}\\
    w2v2-lr-300		& Switchboard & facebook/wav2vec2-large-robust-ft-swbd-300h & \cite{Hsu2021a}\\
    w2v2-lr-960		& Librispeech & facebook/wav2vec2-large-robust-ft-libri-960h & \cite{Hsu2021a}\\
    w2v2-lr-er		& MSP Podcast & audeering/wav2vec2-large-robust-12-ft-emotion-msp-dim & \cite{Wagner2022a,Wagner2022}\\
    w2v2-lr-vad		& MSP Podcast & audeering/wav2vec2-large-robust-12-ft-emotion-msp-dim & \cite{Wagner2022a,Wagner2022}\\
    \hline
    \end{tabular}
    \label{tab:hf}
\end{table*}

\section{Dataset}
This Hume-VB dataset \cite{Cowen2022HumeVB} was used to predict continuous score or class in all four affective vocal bursts tasks. The dataset consists of 59201 samples in 36 hours of recordings. These samples were recorded from 1702 speakers across four countries (China, South Africa, the US, and Venezuela). Each sample was labeled with an integer scale from 1 to 100 (scaled to [0,1] during the experiments) for ten expressed emotions, scores of valence and arousal (in [0, 1]), and the type of vocal bursts. The ten emotion categories are amusement, awe, awkwardness, distress, excitement, fear, horror, sadness, surprise, and triumph. The types of vocal bursts are cry, gasp, groan, grunt, laugh, pant, scream, and other. The original raw audio data were sampled at 48 kHz and resampled to 16 kHz in the experiments of acoustic embedding extraction. The organizer partitioned the data into training, validation, and test sets for each task. Training and validation sets are open with their labels, while the test set is closed without labels. The participants need to send the test set predictions to obtain the performance scores on the test set.

\section{Tasks}
There are four tasks provided in the ACII 2022 affective vocal bursts workshop and competition \cite{Baird2022a}. These tasks are summarized as follows:
\begin{itemize}
    \item The ``High" task is to predict the intensity of 10 emotions.
    \item The ``Two" task is to predict the degree of valence and arousal for given vocal bursts.
    \item The ``Culture" task is to predict the intensity of 40 emotions (10  abovementioned emotions from each culture) as a multioutput regression problem.
    \item The ``Type" task is to predict the type of given vocal bursts. Eight types of vocal bursts in the fourth task are gasp, cry, laugh, scream, groan, grunt, pant, and other.
\end{itemize}

The first three tasks are regression problems evaluated with concordance correlation coefficient (CCC); the fourth task is a classification problem evaluated with unweighted average recall (UAR).

% \section{Baseline Features}
% As the baseline features, we experimented with ComParE \cite{Schuller2013} and eGeMAPS \cite{Eyben} feature sets. These acoustic features are utterance aggregation by calculating high-level statistical functions from the low-level descriptors (extracted per frame). We re-trained these acoustic features on the modified classifiers (see Section \ref{ses:classifier}) to obtain the validation scores. The classifiers are modified from the previous studies \cite{Baird2022a,Atmaja2022c}.

\section{Finetuned wav2vec 2.0}
The wav2vec 2.0 model, including the base (large and robust) and finetuned versions, are the improved version of the previous wav2vec (version 1.0) \cite{Schneider2019}. The wav2vec 2.0 uses a self-supervised approach for generating speech representations \cite{Baevski2020a} instead of unsupervised learning. Given the success of wav2vec 2.0 in speech emotion recognition \cite{Pepino2021}, we evaluated the finetuned versions of wav2vec 2.0 for four affective vocal bursts tasks. These versions are shown in Table \ref{tab:hf}.

For the last model, we extracted two representations of each given vocal burst audio file. The first is a 1024-dims of the hidden states. The second is a 1027-dims of a concatenation of the hidden states and logits (valence, arousal, dominance). These acoustic embeddings were then fed to the regression model for the first three tasks and to the classification model for the last task four ("Type"). The model IDs (in https://huggingface.co) are also given in the Table \ref{tab:hf}.

\section{Classifiers}
\label{ses:classifier}
This study employed a four-layer fully-connected (FC) network as the classifier,
including the output layer. The FC network is a feed-forward
neural network with number of nodes for each layer is 128, 64, and 32,
respectively. The last layer is connected to an output layer. Each FC layer is
connected to layer normalization \cite{Ba2015} and leaky rectified linear unit
(LeakyReLU) activation function. The number of nodes at output layers depends
on the task, i.e., 8 for Type, 2 for Two, 0 for High, and 10 for Culture. The
output layer for regression problems is activated with a sigmoid function.
These classifiers are modifications of the previous studies \cite{Baird2022a,
Atmaja2022b,Atmaja2022c}.

Table \ref{tab:hyperparameters} depicts the hyperparameters of the fully-connected network. The architecture and hyperparameters are the same for all tasks. The learning rate is set to 0.0005, weight decay is set to 0.01, and the maximum number of epochs is 100. For the Type task, an early stopping criterion was set to a patience of 10 epochs. Other tasks are trained without an early stopping criterion of patience. The optimizer is AdamW \cite{Loshchilov2017} with a weight decay of 0.01.

\begin{table}[htbp]
    \caption{Hyperparameters of the classifier}
    \centering\begin{tabular}{l c}
    \hline
    Hyper-parameter & Value \\
    \hline
    Layer & MLP \\
    N\_layers & 4 \\
    Nodes & (128, 64, 32, output\_length) \\
    Normalization & LayerNorm \\
    Layer activation & LeakyReLU \\
    Output activation (regression) & Sigmoid \\
    Optimizer & AdamW \\
    Learning rate & 0.0005 \\
    Weight decay & 0.01 \\
    Number of seeds & 20 \\
    Batch size & 8 \\
    Epochs & 100 \\
    Early Stopping & No (Type: 10 epochs)\\
    Patience delta (Type) & 0.01 \\
    \hline
    \end{tabular}
    \label{tab:hyperparameters}
\end{table}

Three tasks: High, Two, and Culture, minimized CCC loss (since the metric is CCC in a range of [-1, 1]). This loss function is the main difference between this study from the previous studies \cite{Baird2022a,Atmaja2022b,Atmaja2022c} in terms of the model's architecture. Other differences include the use of a different activation function (LayerNorm instead of BatchNorm). The Type task minimized cross-entropy loss. The UAR score for the type is the normalized score in a range of [0, 1].

\section{Open Repository}
The codes to obtain the results are available at https://github.com/bagustris/A-VB2022\_CCC. The repository contains the codes to extract acoustic features, train the classifiers, and evaluate the results (generating predictions). 

\section{Results and Discussion}

% \subsection{Validation benchmark}
Since the labels of the test set are hidden by the organizer of the A-VB 2022, the authors only experimented with the validation set to measure the performance of the evaluated methods (Table \ref{tab:feature}). Two baseline features, eGeMaps \cite{Eyben} and ComParE \cite{Schuller2013}, are compared with five acoustic embeddings. Table \ref{tab:feature} shows the maximum scores from 20 seed numbers evaluations. The scores are CCC for High, Two, and culture; and UAR for Type. Bolds indicate the highest score for each task with `w2v2-lr-er' for Two and culture and `w2v2-lr-vad` for Culture and Type. Both embeddings obtained the same score on the Culture task.

\begin{table}[!htbp]
    \caption{Validation maximum scores from 20 seed numbers on different acoustic features using the same classifier for all tasks (CCC for High, Two, and Culture; UAR for Type)}
    \centering\begin{tabular}{l | c | c c c c}
    \hline
    Feature & Dims. & High & Two & Culture & Type \\
    \hline
    eGeMAPS	    & 88 & 0.4896 & 0.4850 & 0.3880 & 0.3784 \\
    ComParE	    & 6373 & 0.5336 & 0.5122 & 0.4254 & 0.3909 \\
    w2v2-lr	    & 1024 & 0.6292 & 0.6100 & 0.4885 & 0.4734 \\
    w2v2-lr-300	& 1024 & 0.6317 & 0.6285 & 0.5119 & 0.4559 \\
    w2v2-lr-960	& 1024 & 0.5984 & 0.6138 & 0.4961 & 0.4656 \\
    w2v2-lr-er	& 1024 & 0.6521 & \textbf{0.6312} & \textbf{0.5138} & 0.4822 \\
    w2v2-lr-vad	& 1027 & \textbf{0.6523} & 0.6296 & \textbf{0.5138} & \textbf{0.4829} \\
    \hline
    \end{tabular}
    \label{tab:feature}
\end{table}

It is clearly shown in Table \ref{tab:feature} that SSL-based acoustic embeddings outperformed conventional acoustic features on the same classifiers. With smaller dimensions than ComParE dimensions (1024 vs. 6373), most acoustic embeddings obtained better scores. Finetuning wav2vec 2.0 on the affective speech dataset improved scores for tasks; however, comparable scores were also obtained by the large and robust model without and with finetuning on non-affective speech dataset. The latter (w2v2-lr-300) obtained the third position after `w2v2-lr-er' and `w2v2-lr-vad` for the High, Two, and Culture tasks. It is interesting here to see that finetuning on a smaller dataset led to higher performance (w2v2-lr-300 vs. w2v2-lr-960), highlighting the importance of choosing the right data for finetuning. In this case, 300h Switchboard dataset may contain more affective information than 960h Librispeech dataset.

\begin{table*}[hbt!]
    \caption{Average scores on the validation set with standard deviations from 20 seeds (CCC for High, Two, and Culture; UAR for Type)}
    \centering\begin{tabular}{l c c c c}
    \hline
    Feature & High & Two & Culture & Type \\
    \hline
    eGeMAPS	    & 0.485	$\pm$ 0.003	& 0.478	$\pm$ 0.004	& 0.378 $\pm$ 0.003	& 0.352	$\pm$ 0.185 \\
    ComParE	    & 0.527	$\pm$ 0.004	& 0.497	$\pm$ 0.007	& 0.417 $\pm$ 0.003	& 0.375	$\pm$ 0.201 \\
    w2v2-lr	    & 0.620	$\pm$ 0.007	& 0.598	$\pm$ 0.006	& 0.480 $\pm$ 0.007	& 0.448	$\pm$ 0.012 \\
    w2v2-lr-300	& 0.619	$\pm$ 0.006	& 0.618	$\pm$ 0.009	& 0.506 $\pm$ 0.004	& 0.444	$\pm$ 0.006 \\
    w2v2-lr-960	& 0.572	$\pm$ 0.013	& 0.599	$\pm$ 0.006	& 0.491 $\pm$ 0.007	& 0.451	$\pm$ 0.011 \\
    w2v2-lr-er	& \textbf{0.645}	$\pm$ 0.003	& 0.622	$\pm$ 0.006	& 0.507 $\pm$ 0.004	& \textbf{0.462}	$\pm$ 0.013 \\
    w2v2-lr-vad	& 0.644	$\pm$ 0.003	& \textbf{0.623}	$\pm$ 0.004	& \textbf{0.508} $\pm$ 0.004	& 0.461	$\pm$ 0.012 \\
    \hline
    \end{tabular}
    \label{tab:avg}
\end{table*}

On the average performance scores (Table. \ref{tab:avg}), both large-robust model (w2v2-lr) and large-robust model finetuned on 300h Switchboard dataset (w2v3-lr-300) also show competitive results. The `w2v2-lr' obtained a high score for the High task while `w2v2-lr-300' scored competitively on the Two and Culture tasks. Finetuning wav2vec 2.0 large-robust on affective dataset showed superiority on both average and maximum scores evaluations. However, the gap between finetuning on affective (MSP-Podcast) and non-affective (Switchboard) datasets is not large.

When comparing the standard deviations, it was found that the last two embeddings have the most stable results among different seed experiments. Besides the highest scores for all tasks obtained by these two embeddings, these embeddings also performed slightly more stable than others. The evaluation of 20 different seeds for each model is one reason to enable the stability evaluation of the models through their standard deviations. At the same time, the authors could also observe significant differences among specific models.

The authors performed two sample tests (paired samples) to observe if there is any significant difference between the two means of w2v2-lr-er and w2v2-lr-vad. Since the latter only concatenates the former with values of arousal, dominance, and valence, the difference might not be significant (1027-dims vs. 1024-dims). Different 20 seed evaluations enable this statistic test instead of repeating the same seed for different runs. The statistic test shows that the $p-values$ are large for all tasks, indicating that there is no significant difference between the two embeddings ($p-values$ = 0.50, 0.66, 0.51, and 0.80 for High, Two, Culture, and Type tasks, respectively). The degrees of valence, arousal, and dominance are important for obtaining categorial emotions; future research may find different ways of accommodating these values into the acoustic embeddings instead of a simple concatenation performed in this study.

\begin{figure}[htbp]
    \centering
    \includegraphics*[width=0.48\textwidth]{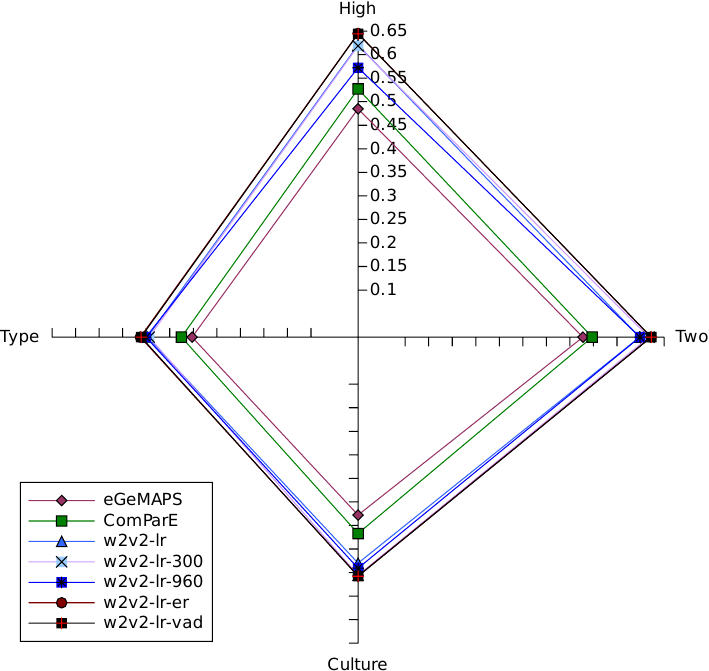}
    \caption{Radar plot of all acoustic features for all tasks.}
    \label{fig:radar}
\end{figure}

Fig. \ref{fig:radar} shows the performance of each model for each task. It could be seen that, in general, one model that performed well in one affective vocal burst task will also perform well in other tasks. In other words, these affective vocal burst tasks are general regardless of the input of acoustic embeddings. One surprising result is that the Type task, which is a classification task, is the most difficult task in this challenge. This is probably because the Type task has a large number of classes (eight) and the number of samples per class is not balanced. This is an interesting finding that should be investigated further; while in most cases, regression is more difficult than classification, in this case, the Type task is more difficult than the other affective vocal bursts tasks.

\begin{figure*}[ht]
    \centering
    \includegraphics*[width=0.7\textwidth]{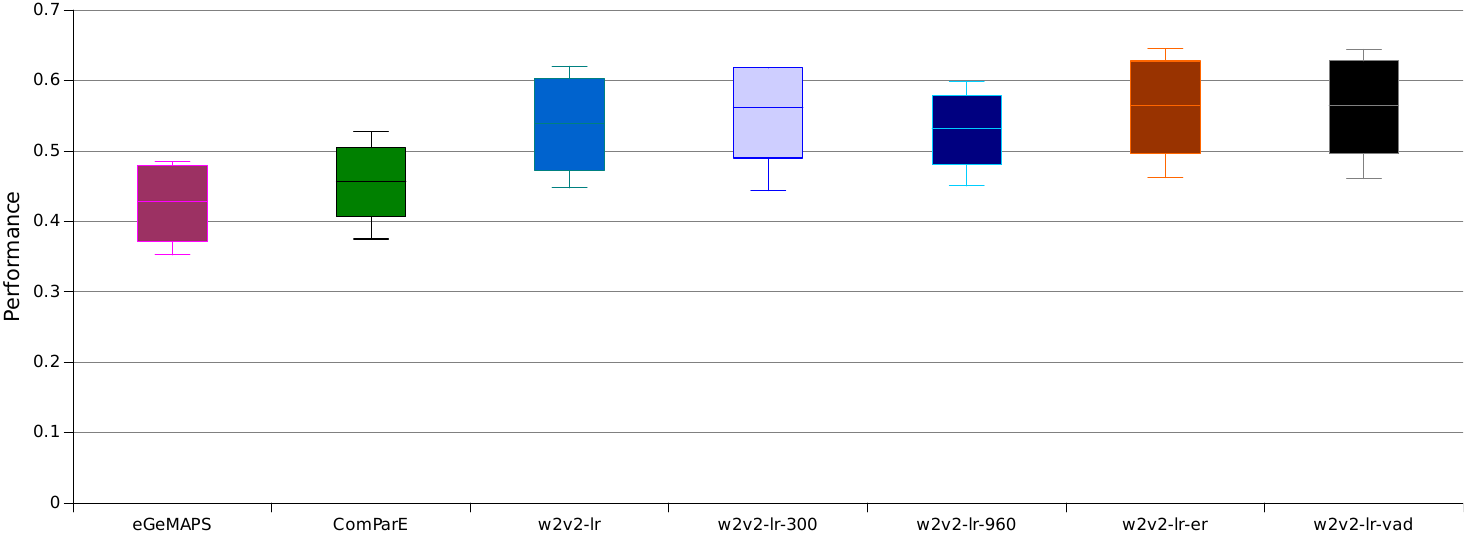}
    \caption{Box plot showing the overall performance (average of four tasks) of all evaluated acoustic features}
    \label{fig:box}
\end{figure*}

Finally, we measure the overall performance for all four tasks by calculating the arithmetic mean across four metrics (three CCCs and a UAR). Since the scores for four tasks are in a range of 0-1 (assumed positive correlation for CCC), it is possible to calculate their average on the same level. The box plot in Fig. \ref{fig:box} shows the overall performance of all evaluated acoustic embeddings. Again, the last two embeddings, w2v2-lr-er and w2v2-lr-vad obtained only slightly better performance than w2v2-lr-300, remaining room for further improvement.

% \begin{figure*}[htbp]
%     \centering
%     % \includegraphics*[width=0.7\textwidth]{average.pdf}
%     \caption{Average scores from 20 seeds for each embedding with their standard deviations}
%     \label{fig:avg}
% \end{figure*}

% \section{Concluding Remark}

% Table \ref{tab:test} shows the test scores of three submitted predictions along with the baseline test results (ComParE \cite{Schuller2013} and eGeMAPS \cite{Eyben}). These three predictions are based on the highest scores on the validation set (Table \ref{tab:feature}). These test scores obtained by three acoustic embeddings are similar to that of validation scores; the only remarkable gap between test and validation is obtained by w2v2-lr-300 (Others) for the Culture task. The overall best score was obtained by w2v2-lr-er; the fusion of w2v2-lr-er with the logits (valence, arousal, dominance) did not improve the scores except for the High task. The w2v2-lr-300 gained comparable high scores to these scores on High and Two tasks, although it was finetuned on non-affective speech dataset.

% say that each participant is given five chances to submit their predictions.
% previous study report two of five with different methods (different embedding and loss function); this study report the rest three of five with different embeddings. Explain the different.

\section*{Acknowledgment}
This technical report is based on results obtained from a project, JPNP20006, commissioned by the New Energy and Industrial Technology Development Organization (NEDO), Japan.

\bibliography{hume}
\bibliographystyle{IEEEtran}

\end{document}